\documentclass[%
reprint,				
 amsmath,amssymb,
 aps,
 prl,
floatfix,
]{revtex4-1}
\usepackage{hyperref}
\hypersetup{
  colorlinks   = true,    
  urlcolor     = blue,    
  linkcolor    = blue,    
  citecolor    = red      
}
\usepackage{xcolor}
\usepackage{graphicx}
\usepackage{dcolumn}
\usepackage{bm}
\usepackage{subcaption}
\def\R{{\bf R}\ }
\def\B{{\bf B}\ }
\def\M{{\bf M}\ }
\def\D{{\bf D}\ }
\begin{document}

\bibliographystyle{unsrt}

\title{Predictive modelling of disease propagation in a mobile, connected community using Cellular Automata}
\author{Ishant Tiwari, Pradeep Sarin}
\author{P. Parmananda}%
\affiliation{%
 Department of Physics, Indian Institute of Technology, Bombay
}%

\date{\today}

\begin{abstract}
We present numerical results obtained from the modelling of a stochastic, highly connected and mobile community. The spread of attributes like health and disease among the community members is simulated using cellular automata on a planar 2 dimensional surface. With remarkably few assumptions, we are able to predict the future course of propagation of such disease as a function of time and the fine-tuning of parameters related to interactions among the automata.
\end{abstract}

\maketitle


The concept of cellular automata placed on a matrix of sites, with rules defining their evolution as a function of time has been very popular since John Conway's proposal \cite{Gardner1970}\cite{White2007,Boccara1993,Pastor-Satorras2015,Bin2019,sinha2017}. When an ensemble is mobile and connected one can observe interesting dynamics such as global synchrony~\cite{sumantra_2010,lavneet_2012} and the simultaneous occurrence of swarming and synchronization~\cite{strogatz_2017}. In this paper we present results involving \emph{mobile} and \emph{connected} cellular automata used to model the spread of a disease in the system. The automata are mobile so at every timestep they jump to neighbouring areas if vacant (we only consider such short distance moves to account for a finite travel velocity). The automata are also `connected' so that each unit has information of the state of all other individuals (through a functional form defined below). The movements are stochastic and strongly depend on the connection relationships. They are also not strictly discrete in their movements, meaning an automaton is allowed to move fractional amounts in the cardinal directions.

\section{Model construction and evolution rules}

\begin{figure}
\includegraphics[width=\linewidth]{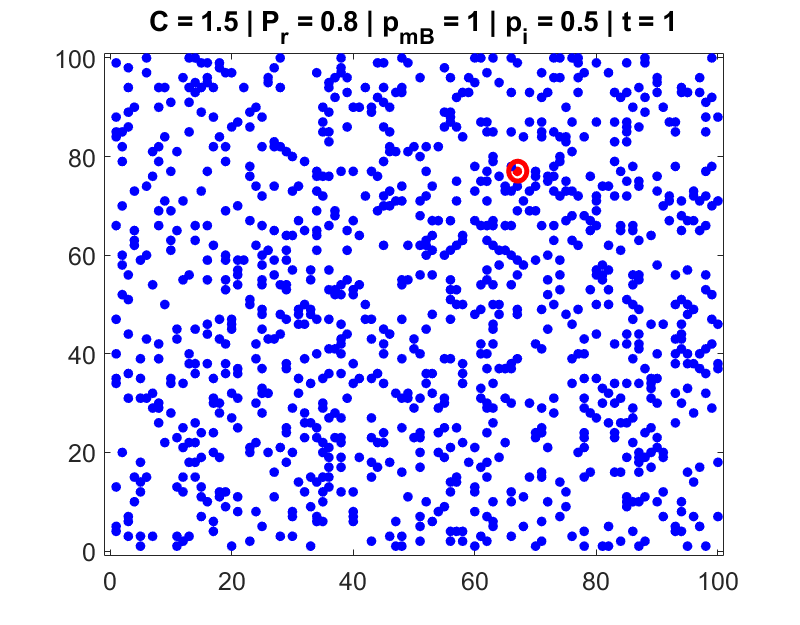}
  \caption{
Model of cellular automata of two types \R and \B (represented by colors Red and Blue respectively). Inital state shown with $N_R=1$ (highlighted with a red circle) and occupancy $\mathrm{O}=0.1$ on a square lattice with $100\times100$ sites}
  \label{fig:initial_state}
\end{figure}

The system initially consists of two types of automata: \B (healthy) and \R (infected). 
Fig \ref{fig:initial_state} shows a visualisation of our system in its initial state on a square lattice in 2 dimensions with $100\times 100$ sites.

The following set of rules define the evolution of the system:
\begin{enumerate}
\item Each automaton can be one of four types \B(healthy),  
 \R(sick,infected), \M(Recovered) or \D(Dead). An automaton changes its type as a functon of time according to rules defined below.

\item The automata are \emph{mobile}, and their mobility is stochastic. At every time step, each automaton moves as directed by a uniformly randomized force in 2-D producing movement by upto 1 unit in the x-y plane, weighted by a force whose direction is determined by its `connectedness' to the other automata of the system as described below.
A combination of these two forces results in movement to points in the 2-D space that are not necessarily on the square lattice sites of the initial condition.
 We \emph{do not} assume periodic boundary conditions: when an automaton reaches the space boundary, its  choice of movement directions is restricted.

\item The system is \emph{connected}. At any time, each automaton has access to information of the type of all automata in the system.
The probability distribution of an automaton's movement direction and magnitude is based on the availability of information about other automata in the following manner:

\begin{enumerate}
\item The probability of a \B jumping to any of the allowed nearby vacant sites is uniform in all directions, weighted with a Yukawa-like force calculated with the potential function:
\begin{equation}
\frac{C}{r}e^{-\frac{r}{L_s}}
\end{equation} 
$r$ is the distance to other automata of type \R in the system, $C$ is a `social distancing factor' and $L_s$ is a length scale. In this sense $L_s$ is the `connection' scale at which the effect of \R is felt over the system and the individual automata can get information about their neighbourhood.
We calculate the repulsive Yukawa `force' exerted on a \B by evaluating its distance $r$ to all the \R in the system.  The vector sum of all these Yukawa forces with magnitude capped at 1 determines the probability of \B jumping in the force direction. Thus \R's far away from a \B hardly affect its jumping direction, while nearby \R's strongly influence the probability of \B to jump away from them. 
With the direction and magnitude of the jump calculated in this manner, the probability of \B actually making that jump is dictated by a movement probability parameter $p_{mB}$.

\item 
We set the probability of \R moving in all directions as equal, with an absolute value of 
$p_{mR}=0.02 p_{mB}$
i.e. \R have less of a tendency to move, compared to \B. This leads to the `herding' the \R's together into relatively stationary clumps and a tendency of the \B's to move away from such clumps as shown in the system snapshot at an intermediate time in Fig \ref{fig:intermediate_state}. The animation of this simulation can be found in the supplementary material (animation.mp4).
\end{enumerate}

\item After every time step, if an \R and \B are within a radius $r=\sqrt{2}$ of each other, $\B\rightarrow\R$ (`infection') occurs with a probability $p_i$. i.e., we allow only person-to-person infection in this model. The background space itself remains pristine: each $(x,y)$ location does not retain a history of its past occupants.

\item Concurrently, we allow two transitions in the state of \R:  \R$\rightarrow$\M (recovery) or \R$\rightarrow$\D (death). 
We assume that once recovered, \M is immune, i.e. \M$\rightarrow$\R is not allowed in this model. The long-term recovery probability of an individual is defined by a parameter $P_r$. The procedure used in the code to translate the long term probability of recovery $P_r$ over $\tau$ timesteps into a recovery probability per time step is explained in the Appendix.

\item After \R$\rightarrow$\M and  \R$\rightarrow$\D transitions, the \M automata move with the same weighted probability distribution set for \B. \D automata are removed from the system.

\end{enumerate}

\begin{figure}
  \includegraphics[width=\linewidth]{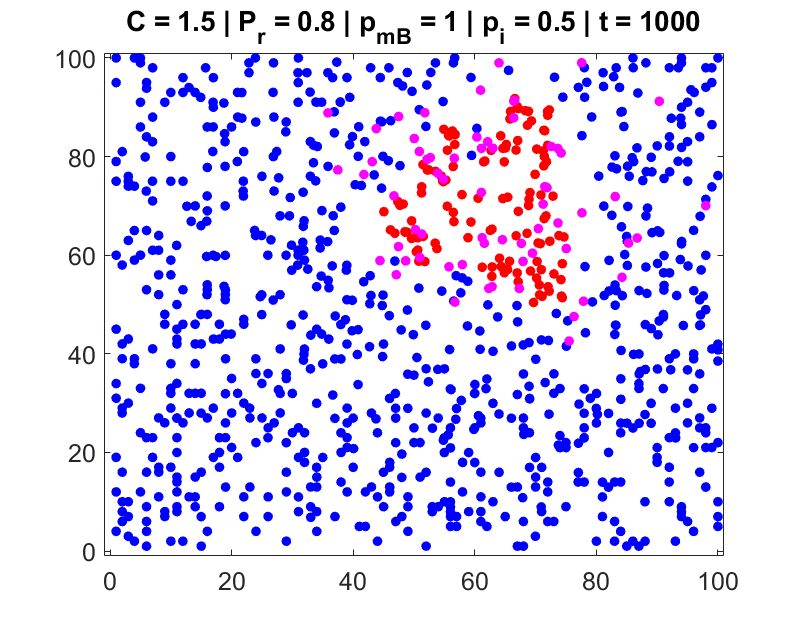}
  \caption{
  Snapshot of system state at an intermediate time $t=1000$ showing distribution of \R (red, infected), \B (blue, healthy) and \M (magenta, recovered) automata, starting from the initial state shown in  Fig \ref{fig:initial_state}
  }
  \label{fig:intermediate_state}
\end{figure}

\section{Initial conditions and model calculations}

We start with the initial condition that $N_0^0$ automata are spread uniformly on $100\times100$ regularly spaced grid. As determined by the movement rules above, the location of automata immediately spreads out in continuous $(x,y)$ because of the Yukawa-like social distancing force. The occupancy factor $\mathcal{O}=N_0^0/10^4$ is set to $0.1$ for computational speed.
 We set $N_0^0=1000, N_R=1$ and $N_B = 999$ as shown in Fig~\ref{fig:initial_state} \\
 The site for placement of the lone $N_R$ is chosen at random.

The main performance parameters dynamically tracked in our model calculation are
$N_R(t), N_B(t), N_M(t), N_0(t)$ the number of infected(\R), healthy(\B), recovered(\M) and total ($N_0=N_0^0 - N_\D$) automata respectively at every time step. Of these, perhaps the most important parameter is $N_R^{max}$, the maximum number of \R reached.

\begin{figure}
\begin{subfigure}{.5\linewidth}
\centering	
  \includegraphics[width=\linewidth]{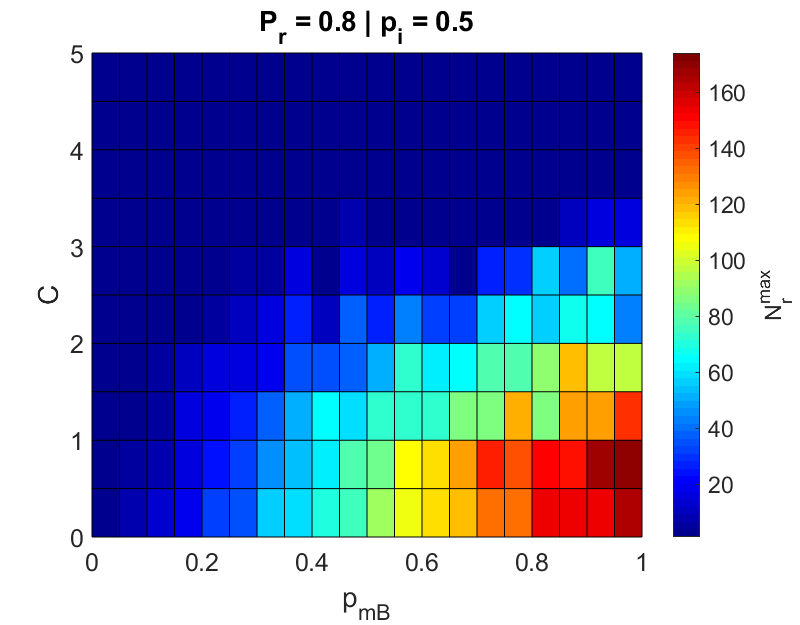}
  \subcaption{$(C,p_{mB})$}
  \label{fig:C_pmb_heatmap}
\end{subfigure}
\begin{subfigure}{.5\linewidth}
\centering
  \includegraphics[width=\linewidth]{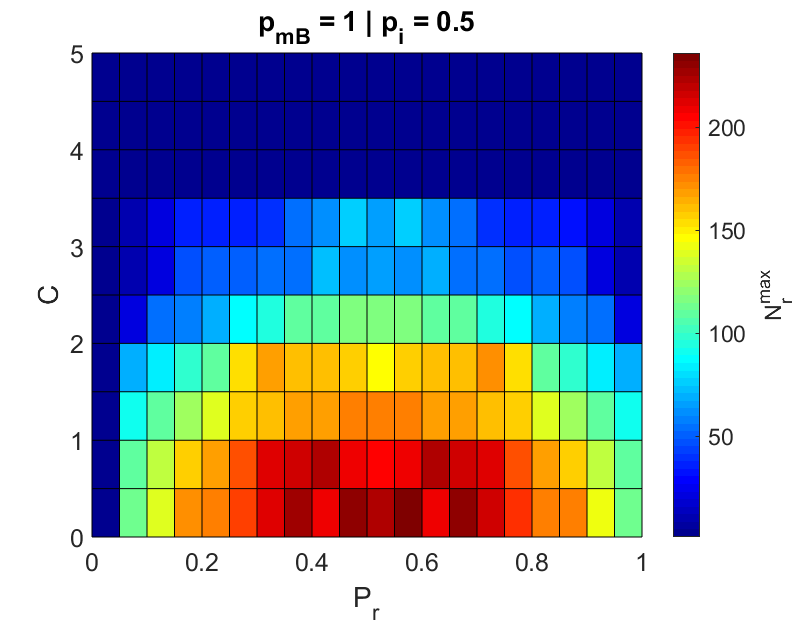}
	\subcaption{$(C,P_{r})$}
  \label{fig:C_Pr_heatmap}
\end{subfigure}
\begin{subfigure}{.5\linewidth}
\centering
 \includegraphics[width=\linewidth]{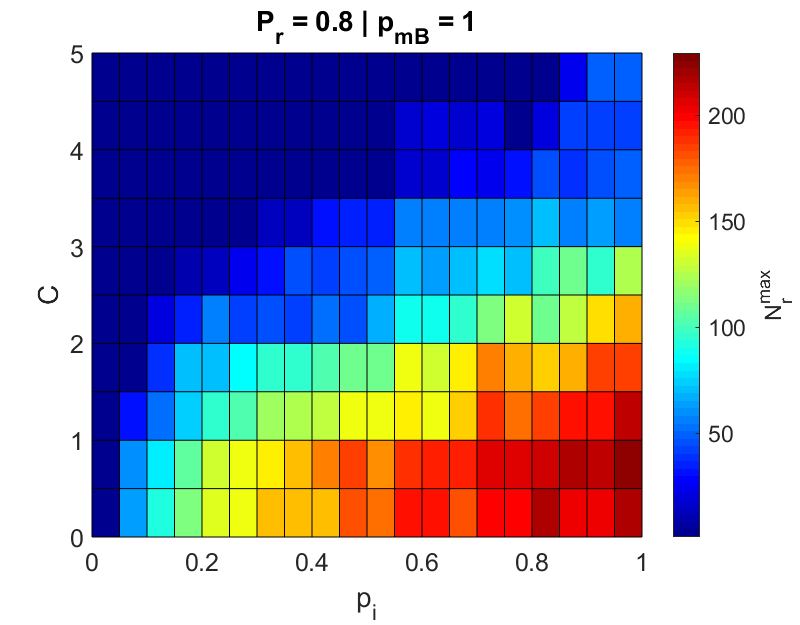}
	\subcaption{$(C,p_{i})$}
  \label{fig:C_pi_heatmap}
\end{subfigure}
  \caption
  {
  	Phase diagram of the maximum number of sick individuals $N_R^{max}$ reached as a function of 
  $(C,p_{mB})$~(\ref{fig:C_pmb_heatmap}), $(C, P_r)$~(\ref{fig:C_Pr_heatmap}) and 
  $(C, p_i)$~(\ref{fig:C_pi_heatmap})
  }
\end{figure}

The time evolution of the system is explored with a phase space of four control parameters $(C, p_{mB}, P_r, p_i)$. 
\begin{enumerate}
\item Social distancing factor $C$ is used to calculate the weighting factor $\frac{C}{r}e^{-\frac{r}{L_s}}$ in deciding the probabilistic direction of movement of \B. $L_s=10$ is fixed for this study. $C=0$ implies ignorance of \R in deciding mobility direction of \B. Increasing values of $C$ cover a large circle of connectedness around \B over which proximity of \R's affects  the direction of \B mobility.

\item The effect of `quarantine' on \B, i.e. to reduce the movement probability $p_{mB}$ of \B. Note that we simply reduce $p_{mB}$ for the entire system evolution, not for a subset of the automata or intermittent periods. 

\item The recovery probability $P_r$ which is somewhat correlated with $N_R^{max}$ but can be also adjusted by ramping up the treatment capacity in the system as a function of time - we have not explored changing $P_r$ with $t$ in this calculation.

\item Infection probability $p_i$ which is often related to a demographic mapping of the population. Since we take all the \B automata as equal, we have set $p_i$ to be the same for all \B and explored the phase space of $(C,p_i)$ which gives a much finer mapping of the effect of $C$ and $p_i$ separately on the system dynamics.
\end{enumerate}

We consider $P_r, p_{mB}, p_i$ to be orthogonal to each other, and study the system evolution in the 
$(C,P_r), (C,p_{mB}), (C,p_i)$ phase space.

\begin{figure}
  \centering
   \setbox1=\hbox{\includegraphics[width=\linewidth]{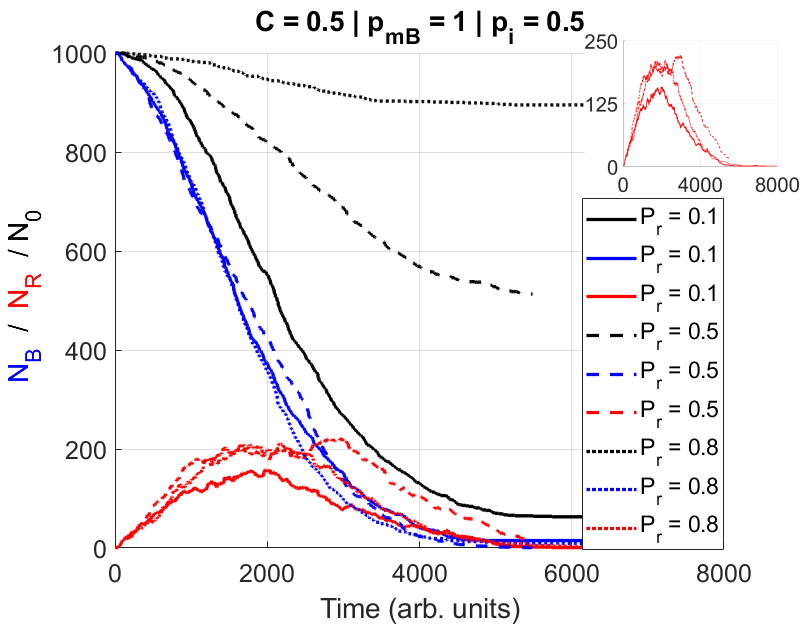}}
  \includegraphics[width=\linewidth]{fig_N_r_t_p_mB.png}
  \caption{
   Time trends of population numbers 
  	for the worst case scenario shown in Fig \ref{fig:C_pmb_heatmap}.
   $N_R^{max}$ is higher for $P_r = 0.5$, compared to that for $P_r = 0.1$, $0.8$.
   Inset shows zoomed in evolution of the sick population $N_R(t)$.
  	}  
  \label{fig:N_r_t_p_mB}
\end{figure}

\begin{figure}
\centering
  \setbox1=\hbox{\includegraphics[width=\linewidth]{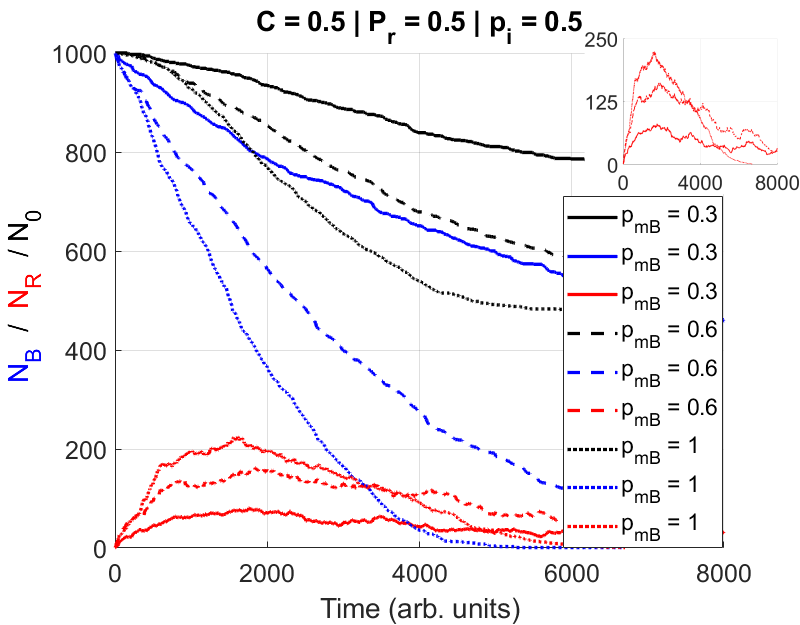}}
	\includegraphics[width=\linewidth]{fig_N_R_t_p_r.png}
  \caption{
  	Time trends of population numbers 
    for the worst case scenario shown in Fig \ref{fig:C_Pr_heatmap}.
   Time of reaching $N_R^{max}$ is \textit{independent} of $p_{mB}$ although
   $N_R^{max}$ increases with increasing $p_{mB}$.
Inset shows zoomed in evolution of the sick population $N_R(t)$.
  }
  \label{fig:N_r_t_p_r}
\end{figure}

\begin{figure}
\centering
    \setbox1=\hbox{\includegraphics[width=\linewidth]{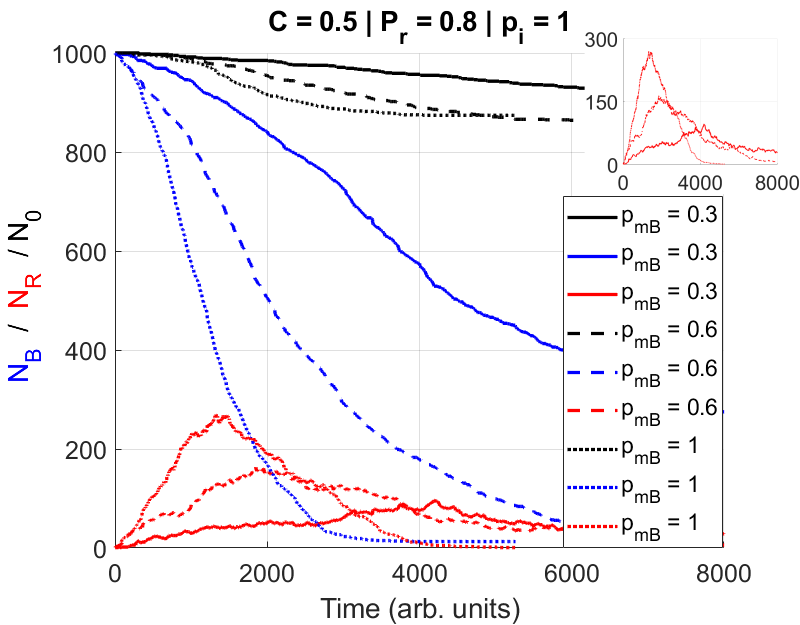}}
	\includegraphics[width=\linewidth]{fig_N_R_t_p_i.png}
  \caption{
 	Time trends of population numbers
    for the worst case scenario shown in Fig \ref{fig:C_pi_heatmap}.
    Lower mobility ($p_{mB}$) delays the time of arrival, and magnitude of $N_R^{max}$.
Inset shows zoomed in evolution of the sick population $N_R(t)$.
  	}
  \label{fig:N_r_t_p_i}
\end{figure}

\section{Numerical results}

We have run the cellular automata model using the initial conditions and rules listed in the previous section for 10 trials, each with a different random number seed.
Fig~\ref{fig:C_pmb_heatmap} shows the phase diagram of maximum \R population $N_R^{max}$ reached during the system evolution for variation in each parameter pair $(C,P_r), (C,p_{mB}), (C,p_i)$. The $N_R^{max}$ displayed in Fig~\ref{fig:C_pmb_heatmap} is averaged over 10 trials. We find in all cases that keeping the social distancing factor $C>2.5$ is invariably able to mitigate $N_R^{max}$ to less than 10\% of the initial population $N_0^0$, regardless of the other parameters. 

However, it is interesting also to look at the time evolution of the sub-populations, especially 
$N_R(t)$. Figs~\ref{fig:N_r_t_p_mB}~-~\ref{fig:N_r_t_p_i} show the {\bf worst} case scenarios in any one of the trials. We pick the highest $N_R^{max}$ cell in the phase map of Fig~\ref{fig:C_pmb_heatmap} and show its $N_R(t)$ for variation in orthogonal parameter $P_r$ in Fig~\ref{fig:N_r_t_p_mB}. The same is done  for 
Fig~\ref{fig:C_Pr_heatmap}~-~Fig \ref{fig:N_r_t_p_r}
and 
 Fig~\ref{fig:C_pi_heatmap}~-~Fig \ref{fig:N_r_t_p_i}.
It is interesting to note in Fig~\ref{fig:N_r_t_p_mB} that the $N_R^{max}$ achieved for a higher recovery parameter ($P_r = 0.5$) is higher (therefore, worse) than that for $P_r = 0.1$. This can be explained by considering the fact that any increase in $N_R$ occurs via the transition of \B $\rightarrow$ \R, while a decrease in $N_R$ is possible via two mechanisms, namely, \R $\rightarrow$ \M and \R $\rightarrow$ \D. Since we consider $P_r + P_d = 1$ in our simulations, for $P_r = 0.1$, the value of $P_d = 0.9$. Therefore, even though the probability of removal of \R via \R $\rightarrow$ \M transition is low, the probability of \R $\rightarrow$ \D is high. One can therefore infer that both a higher mortality rate ($P_r = 0.1, P_d=0.9$) and higher recovery rate ($P_r = 0.8, P_d=0.2$) can result in a lower number of simultaneously sick individuals (and hence $N_R^{max}$) as seen in Fig~\ref{fig:N_r_t_p_mB}. 

In Fig~\ref{fig:N_r_t_p_r}, one can see that a higher mobility (higher $p_{mB}$) causes $N_R$ to rise at a faster rate. This is evident from the observation that a higher value of $N_R^{max}$ is obtained for a higher $p_{mB}$ value, but at approximately the same time. Subsequently, in Fig~\ref{fig:N_r_t_p_i}, since the probability of `infection' $p_i = 1$, the effect of mobility is more drastically observed (i.e., higher $p_{mB}$ causes higher $N_R^{max}$). From this one can infer that a highly contagious disease will be more sensitive to the strictness of quarantine that individuals observe. 

The small fluctuations in population numbers of infected($N_R(t)$), healthy($N_B(t$)), and total ($N_0(t)$) reflect the inherent stochasticity of our calculations, much like real data from human populations\cite{owidcoronavirus} but not usually reflected in system dynamics models of epidemiology. These fluctuations depend on the initial conditions used in our simulation which are determined by  the seeds of the random number generator.

\section{Summary}

We have developed a general flexible framework for modeling the propagation of attributes in a stochastic system of mobile, connected automata. Using simple evolution rules, we have obtained interesting results on the evolution of a disease epidemic in one such closed system. Considering the generic nature of the model that we have considered, we emphasized that this work can find applications in a plethora of different settings in addition to the epidemiological modeling performed here. A few possible avenues which could be explored with minimal tweaking of the present model include network theory, transportation modeling, and ecological modeling of the interaction of various species in an ecosystem.\\
In the future, one could study the effect of a distribution of parameter values present in the population, instead of just one universal parameter value applicable for every entity. This would better model a mixture of population with some agents more susceptible than others to an infection, or a sub-section of population having co-morbidities which can increase their mortality rate after infection. Another possibility is to create a time variation of parameters emulating effects like the slow rise in complacency an entity might experience in trying to avoid getting infected.   \\\\

\section{Acknowledgments}
We are grateful for support provided by DST grants EMR/2016/000275 (P.P.),  SR/MF/PS-02/20014-IITB (P.S) and CSIR (I.T.). This work was supported by the Department of Physics, IIT Bombay.

\section{Appendix}
We define $P_r$ and $P_d$ as the \textit{integrated} probability over a timescale $\tau$ that an entity will recover or die after being infected. Note that we have set $\tau=1000$ in our calculations - this includes the possibility that over that time period, the entity continues propagating its sickness to others without itself showing any symptoms of sickness.

In our model, we have enforced $P_r + P_d=1$.
Let the per iteration probability of recovery be $p_r$, and similarly of death be $p_d$. In a single iteration,
$p_r + p_d \neq 1$ because there is an additional probability $p_s$ that an entity will remain sick. Therefore, in a single time step: \mbox{ $p_r + p_d + p_s =1$}
The probability of an entity \textit{NOT} recovering in a time step is $(1-p_r)$. 

In $\tau$ iterations, the probability of an entity not recovering is $(1-p_r)^\tau$.
Assuming behavior in successive time steps is independent after infection occurs until either recovery or death: 

\begin{eqnarray}
(1-p_r)^\tau = (1-P_r) 					\\
(1-p_r) = (1-P_r)^\frac{1}{\tau} 		\\
p_r = 1 - (1-P_r)^\frac{1}{\tau} 
\end{eqnarray}

\end{document}